\documentclass[aps,prb,twocolumn]{revtex4} 

\usepackage{graphicx}
\usepackage{dcolumn}
\usepackage{bm}
\usepackage{amsmath}  

\newcommand{\comment}[1]{}

\begin{document}
\renewcommand{\theequation}{\arabic{section}.\arabic{equation}}

\title{Double Fountain  Effect in Superfluid Helium}


\author{Phil Attard}
\affiliation{ {\tt phil.attard1@gmail.com}}


\begin{abstract}
A double fountain pressure model is used to analyze
the recent measurements of Yu and Luo (arXiv2211.02236v4)
of superfluid $^4$He flow between two chambers
held at different temperatures via two superleaks
and an intervening third chamber that spontaneously achieves
a  temperature higher than both fixed temperatures.
The physical origin of the increased temperature in the intervening chamber
is attributed to the balance between the rate of mechanical energy
deposited by superfluid transport
and the rate of convection back to the lower temperature chambers.
An equation is given for the pressure of the third chamber,
the measurement of which would confirm or refute the theory.
\end{abstract}


\maketitle

%
%

\renewcommand{\theequation}{\arabic{equation}}

\begin{figure}[t!]
\centerline{ \resizebox{7cm}{!}{ \includegraphics*{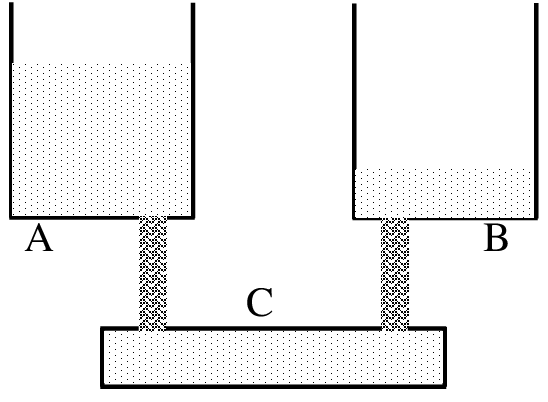} } }
\caption{\label{Fig:ABC}
Two chambers containing saturated $^4$He at fixed temperatures
$T_A < T_B < T_\lambda$
and connected by superleaks to a closed container C.
}
\end{figure}

Recently Yu and Luo  (2022)
carried out  measurements on superfluid $^4$He
below the $\lambda$  transition temperature $T_\lambda$
using the experimental arrangement depicted in figure~\ref{Fig:ABC}.
Starting empty,
the high temperature chamber B gradually fills
at a steady rate over many hours.
After about 2 hours chamber C attains a steady temperature $T_C$
that is higher than the fixed temperatures
of either of the two other chambers,
but still below the $\lambda$-transition temperature,
$T_\lambda > T_C > T_B > T_A$.

The high temperature induced in chamber C
by the superfluid flow  is at first sight surprising.
Yu and Luo  (2022) conclude:
\begin{center}
\parbox{7.5cm}{
``The two-fluid model proposes that a super flow of $^4$He
carries no thermal energy\ldots This experimental result
directly contradicts
the pivotal hypothesis of the two-fluid model''
(Yu and Luo 2022 page 5).
}\end{center}
And also
\begin{center}
\parbox{7.5cm}{
``The two-fluid model postulates the existence of a super
fluid component that possesses an exotic characteristic of zero entropy\ldots
the zero entropy assumption requires this temperature to be absolute zero''
(Yu and Luo 2022 page 3).
}\end{center}
To resolve these problems
Yu and Luo (2022) propose a replacement for the two-fluid model,
namely that low-lying energy levels occur in particular groups
that are thermally populated by $^4$He atoms.

The claim that the measurements refute the accepted model
of superfluidity merits close scrutiny.
The interpretation of the conventional two-fluid model
by Yu and Luo (2022) is not without foundation.
F. London (1938) explained superfluidity and the $\lambda$-transition
as Bose-Einstein condensation into  the ground energy state,
as Einstein (1925) had explicitly proposed (Balibar 2014).
Tisza (1938) explained superfluid hydrodynamics
by postulating that helium~II had zero entropy.
Landau's (1941) phonon-roton theory
focusses on the ground state for helium~II,
(and solely the first excited state for helium~I).
H. London (1939)
derived the fountain pressure equation,
for which there is overwhelming quantitative experimental evidence
(Donnelly  and  Barenghi  1998),
by asserting that condensed bosons have zero entropy.

In contrast, I believe that the derivation of the fountain pressure equation
by H. London (1939) is flawed,
although the equation itself is correct,
and that in fact the equation implies that superfluid flow
is at constant entropy,
not zero entropy (Attard 2022, 2023a, 2023b).
Also I believe that Bose-Einstein condensation
is into multiple low-lying momentum states,
not the ground state (Attard 2023a, 2023c).
In this paper I show how
the Bose-Einstein condensation, two-fluid model of superfluidity,
modified with these two ideas,
explains the  temperature increase observed by Yu and Luo (2022).

\comment{ 
There are three reasons why it is worthwhile to analyze
the experimental set up developed by Yu and Luo (2022).
First, an alternative model of the measurements
with plausible (and testable) predictions
restores faith in the conventional understanding of superfluidity
in terms of Bose-Einstein condensation
and  removes the need to resort to more exotic theories.
Second, it confirms that
the condensed bosons involved in superfluid flow
have both entropy and energy,
it emphasizes the driving force for superfluid flow,
and it explains how the temperature of the third chamber increases.
And third, the analysis shows
how the experimental arrangement can be used
to measure an important superfluid transport coefficient
that appears otherwise unattainable.
} 

The experimental arrangement (figure~\ref{Fig:ABC})
is modeled as two fountain pressure systems
with a common high temperature, high pressure chamber C.
As mentioned,
the original fountain pressure equation (H. London 1939)
has stood the test of time,
although I have a different derivation and interpretation
(Attard 2022, 2023a).
I find that permutation entropy is what drives Bose-Einstein condensation,
and that the fountain pressure equations
imply that condensed bosons
are driven to minimize the energy at constant entropy,
which leads to equality of chemical potential (Attard 2022, 2023a).
This is thermodynamically equivalent to the  H. London (1939)
formula for the fountain pressure (Attard 2022, 2023a).

The fact that the present system is in a steady state with continuous flow
from A through C to B does not fundamentally affect the analysis
since for slow changes one can  invoke local thermodynamic equilibrium.
It is standard for the fountain pressure to be measured
with steady flow in the superleak  (Keller and Hammel 1960).
What drives the fountain pressure,
and superfluid flow more generally,
is the minimization of the energy at constant entropy,
which is equivalent to chemical potential equality
between connected superfluid regions (Attard 2022, 2023a).
In the present steady state case
with chambers A and B held at different temperatures, $T_B > T_A$,
equality is not possible because
the chemical potential decreases
with increasing temperature along the saturation curve,
$ \mu_B^\mathrm{sat} < \mu_A^\mathrm{sat}$
(Donnelly  and  Barenghi  1998).
Instead the most that can be achieved in the steady state
is for the chemical potential in chamber C
to be the average of that in the fixed temperature chambers,
\begin{equation}
\mu_C = \frac{1}{2} [ \mu_A^\mathrm{sat} + \mu_B^\mathrm{sat} ].
\end{equation}
With this the difference in chemical potential across each superleak  is
$ \Delta_\mu = \mu_A^\mathrm{sat} - \mu_C
= \mu_C - \mu_B^\mathrm{sat} > 0$.
Below I show that if the superfluid number flux
is proportional to the difference in chemical potentials,
then for identical superleaks
this result ensures mass conservation.

For an incompressible liquid,
the change in pressure equals the number density
times the change in chemical potential (Attard 2002).
Hence the preceding result gives the pressure in chamber C as
\begin{eqnarray}
p_C
& = &
 p_C^\mathrm{sat} + \rho_C^\mathrm{sat} [ \mu_C -\mu_C^\mathrm{sat}]
\nonumber \\ & = &
 p_C^\mathrm{sat} +  \frac{\rho_C^\mathrm{sat}}{2}
 [ \mu_A^\mathrm{sat} + \mu_B^\mathrm{sat}  - 2\mu_C^\mathrm{sat} ].
\end{eqnarray}
Measured values as a function of temperature
for the various quantities on the right hand side have been tabulated
for $^4$He (Donnelly  and  Barenghi   1998).
If one measures the temperature $T_C$ then this gives the pressure  $p_C$.
Measuring both $T_C$ and $p_C$ would
confirm or refute the present  double fountain pressure model
and analysis of the experimental arrangement.

A second equation explains
the elevated temperature of the closed intermediate chamber.
Since superfluid flow is driven
to equalize the chemical potential (Attard 2022, 2023a),
the simplest assumption is that in the steady state
the number flux in each superleak is linearly proportional
to the chemical potential difference across it,
\begin{equation}
J_{N,AC}
= K_{AC} [ \mu_A^\mathrm{sat} - \mu_C ]
= K_{AC} \Delta_\mu .
\end{equation}
The superfluid transport coefficient depends on
the material properties  of the superleak,
likely scaling with the cross-sectional area
while being independent of the length.
The validity of this assumed linear form
should be checked by actual measurement.
Similarly
\begin{equation}
J_{N,CB}
= K_{CB} [ \mu_C - \mu_B^\mathrm{sat} ]
= K_{CB} \Delta_\mu .
\end{equation}
In the steady state, mass conservation gives $J_{N,AC} = J_{N,CB}$.
For identical superleaks,
$K_{AC}$ = $K_{CB}$,
and so these equations confirm that the chemical potential difference
must be the same across the two superleaks.

The superfluid  flows at constant entropy
(Attard 2022, 2023a, 2023b).
The fountain pressure equation  (H. London 1939)
minimizes the energy at constant entropy
(Attard 2022, 2023a).
Since
$\partial E(S,V,N)/\partial N = \mu$
(Attard 2002),
the rate of energy transport by superfluid flow
is just the chemical potential times the number flux.
Hence the rate of energy change of chamber $C$
due to superfluid flow through it is
\begin{eqnarray}
\dot E_C^\mathrm{sf}
& = &
[ \mu_A^\mathrm{sat} J_{N,AC} - \mu_C J_{N,CB}]
\nonumber \\ & = &
[ \mu_A^\mathrm{sat}  - \mu_C  ]J_{N}
\nonumber \\ & = &
K \Delta_\mu^2.
\end{eqnarray}
This is positive irrespective of which chamber  has the higher temperature.
This assumes that there is no gradient in chemical potential
within the superleaks,
so that there is a step change in chemical potential at their exits.
This result shows that superfluid flow carries energy,
and it explains how chamber C is heated by that flow.

In the steady state  this superfluid energy flux into the chamber
must be equal and opposite to the heat flow from the chamber
to the two fixed temperature chambers A and B.
The heat flux in helium~II
has been measured (F. London and Zilsel 1948, Keller and Hammel 1960),
including in powdered superleaks (Schmidt and Wiechert  1979).
The rate of change of the energy in chamber $C$ due to conduction
via the walls, powder, and liquid of the superleaks
is proportional to  the  temperature gradients,
\begin{eqnarray}
\dot E_C^\mathrm{cond}
& = &
\Lambda_{AC} L_{AC}^{-1} [ T_C^{-1}-T_A^{-1}]
+
\Lambda_{AB} L_{AB}^{-1} [ T_C^{-1}-T_B^{-1}]
\nonumber \\ & = &
\Lambda  L^{-1} [ 2 T_C^{-1}-T_A^{-1} -T_B^{-1}]
\nonumber \\ & \equiv &
\Lambda  L^{-1} \Delta_T^\mathrm{tot}.
\end{eqnarray}
This is just Fourier's law (in inverse temperature),
with $\Lambda$ being the effective thermal conductivity,
and $L$ the length of the superleak.
If the temperature of C is greater than the fixed temperatures,
$T_C > T_B > T_A$,
then $\Delta_T^\mathrm{tot} < 0$,
and energy is conducted out of chamber C.
Evidently and obviously, the larger $T_C$,
the greater the rate of energy loss by conduction.
If conduction is the dominant mechanism for the heat back-flow,
then the chamber temperature $T_C$
is determined by the steady state condition,
$\dot E_C^\mathrm{cond} + \dot E_C^\mathrm{sf} = 0$.

In fountain pressure measurements
there is viscous flow from the high pressure chamber
through the connecting capillary, frit, or superleak
(Keller and Hammel 1960).
The viscous number flux from chamber  C
should be linearly proportional to the
sum of the pressure gradients,
$ \Delta_p^\mathrm{tot}/L
\equiv [ 2 p_C - p_A^\mathrm{sat} -p_B^\mathrm{sat}]/L $.
Hence the convective rate of energy change  scales as
\begin{equation}
\dot E_C^\mathrm{conv}
\propto -\Delta_p^\mathrm{tot} h_C
\approx -\Delta_p^\mathrm{tot} h_C^\mathrm{sat} ,
\end{equation}
where $h$ is the enthalpy per particle,
which is taken at saturation to make use of readily available data.
If convective heat flow dominates the  heat back-flow,
then $T_C$
is determined by the steady state condition,
$\dot E_C^\mathrm{conv} + \dot E_C^\mathrm{sf} = 0$.
Presumably radiation losses  are negligible.

\begin{table}[t!]
\caption{ \label{Tab:t1}
Measured temperatures (Yu and Luo 2022),
and calculated quantities.
}
\begin{center}
\begin{tabular}{c c c c c c  }
\hline\noalign{\smallskip}
$T_A$ & $T_B$ & $T_C$ & $p_C$ &
$\displaystyle \frac{- \Delta_\mu^2 }{\Delta_T^\mathrm{tot}} $  &
$\displaystyle \frac{ \Delta_\mu^2
}{ h_C^\mathrm{sat}\Delta_p^\mathrm{tot}}$ \\
(K) & (K) & (K) & (kPa) & (--) & (--)   \\
\hline
 1.500(4) & 1.700(4) & 1.847(1) & 15.5 & 0.9 & 18 \\ 
 1.600(4) & 1.800(4) & 1.927(1) & 18.9 & 2.1 & 23 \\ 
 1.600(4) & 1.900(4) & 2.014(1) & 25.9 & 6.2 & 42 \\
\hline
\end{tabular}
\end{center}
\end{table}


In table~\ref{Tab:t1}
the measured temperatures (Yu and Luo 2022)
are used to test these results.
The saturated chemical potentials and enthalpies are derived from data
given by Donnelly  and  Barenghi (1998),
corrected as explained by Attard (2022, 2023a).
The predicted pressure $p_C$ is substantially higher
than the saturated vapor pressures
(eg.\ $p^\mathrm{sat}(1.5\,K) = 0.47$\,kPa
and  $p^\mathrm{sat}(2.0\,K) = 3.13$\,kPa)
(Donnelly  and  Barenghi 1998).
As mentioned, comparison of the calculated and measured pressure
would test the present theory.

According to the present theory,
if conduction dominates,
the ratio $-\Delta_\mu^2/\Delta_T^\mathrm{tot}$
should be positive and constant in any one series of measurements.
If convection dominates,
$\Delta_\mu^2/( h_C^\mathrm{sat}\Delta_p^\mathrm{tot})$
should be positive and constant.
In both cases in table~\ref{Tab:t1} the energy flux ratio is positive.
Over the series of measurements it varies by about a factor of seven
for conductive heat flow,
and by about a factor of two for convective heat flow.
These results suggest
that it is mainly heat transported by viscous flow
that  counters the superfluid energy flux
and stabilizes the steady state temperature $T_C$.
Further measurements are required to quantitatively clarify the situation.


In conclusion, the double fountain pressure model
provides a basis to analyze the experimental arrangement
of Yu and Luo (2022).
The results confirm that superfluid flow carries both entropy and energy
and that it is driven to minimize energy at constant entropy
(Attard 2022, 2023a, 2023b).
This gives a physical mechanism for the temperature increase
in the closed chamber,
and it reconciles the experimental measurements
with the (modified) Bose-Einstein condensation,
two-fluid model of superfluidity.
The experimental design of  Yu and Luo (2022)
might provide a quantitative measurement method
for the superfluid transport coefficient.

\section*{References}


\begin{list}{}{\itemindent=-0.5cm \parsep=.5mm \itemsep=.5mm}

\item 
Attard  P 2002
\emph{Thermodynamics and Statistical Mechanics:
Equilibrium by Entropy Maximisation}
(London: Academic Press)

\item  
Attard P (2022)
Further On the Fountain Effect in Superfluid Helium
arXiv:2210.06666 (2022)

\item 
Attard  P 2023a
\emph{Entropy beyond the second law.
Thermodynamics and statistical mechanics
for equilibrium, non-equilibrium, classical, and quantum systems}
(Bristol: IOP Publishing, 2nd edition)

\item 
Attard  P 2023b
Quantum Stochastic Molecular Dynamics Simulations
of the Viscosity of Superfluid Helium
arXiv:2306.07538 (2023b)

\item 
Attard P 2023c
The Paradox of Bose-Einstein Condensation
arXiv:2307.11743

\item 
Balibar S (2014)
Superfluidity: How Quantum Mechanics Became Visible.
In:
Gavroglu, K. (eds)
\emph{History of Artificial Cold, Scientific,
Technological and Cultural Issues.}
Boston Studies in the Philosophy
and History of Science, {\bf 299}
(Dordrecht: Springer)

\item 
Donnelly R J and  Barenghi C F 1998
The observed properties of liquid Helium at the saturated vapor pressure
\emph{J.\ Phys.\ Chem.\ Ref.\ Data} {\bf 27} 1217

\item
Einstein A 1925
Letter to Paul Ehrenfest (Balibar 2014)

\item  
Keller W E  and Hammel (Jr) E F 1960
Heat conduction and fountain pressure in liquid He~II
\emph{Annals of Physics} {\bf 10}  202

\item  
London F 1938
The $\lambda$-Phenomenon of Liquid Helium and the Bose-Einstein Degeneracy
\emph{Nature} {\bf 141} 643

\item 
London H 1939
Thermodynamics of the thermomechanical effect of liquid He II
\emph{Proc.\ Roy.\ Soc.}\ {\bf  A171} 484

\item  
London F  and  Zilsel P R 1948
Heat transfer in liquid helium II by internal convection
\emph{Phys.\ Rev.}\ {\bf 74} 1148

\item
Schmidt R and Wiechert H 1979
Heat Transport of Helium II in Restricted Geometries
\emph{Z. Physik B} {\bf 36} 1

\item  
Tisza L 1938
Transport phenomena in helium II
\emph{Nature} {\bf 141} 913

\item 
Landau L D 1941
Theory of the superfluidity of helium II
\emph{Phys.\ Rev.}\ {\bf 60} 356

\item
Yu Y and Luo  H (2022)
Microscopic Picture of Superfluid $^4$He
arXiv2211.02236v4

\end{list}


\end{document}